\documentclass[twocolumn,showpacs,preprintnumbers,amsmath,amssymb,prl,superscriptaddress]{revtex4-1}
\usepackage{graphicx}% Include figure files
\usepackage{dcolumn}% Align table columns on decimal point
\usepackage{bm}% bold math
\usepackage{epsfig}
\usepackage{natbib}
\usepackage{color}

\newcommand{\me}{m_\text{0}} %electron free mass
\newcommand{\mb}{m_\text{b}} %electron band mass
\newcommand{\lB}{l_\text{B}} % magnetic length
\newcommand{\rs}{r_\text{s}} % Wigner Seitz radius
\newcommand{\ttr}{\tau_\text{tr}} % transport scattering time in zero field
\newcommand{\tq}{\tau_\text{q}}  % quantum scattering time
\newcommand{\Rxx}{R_\text{xx}}  % quantum scattering time
\begin{document}
%\preprint{PRL/CF mass}

\title{Temperature dependent magnetotransport around $\nu$= 1/2 in ZnO heterostructures}
\author{D.~Maryenko}
\email{maryenko@riken.jp}
\affiliation{Correlated Electron Research Group (CERG), RIKEN Advanced Science Institute, Wako 351-0198, Japan}
\author{J.~Falson}
\affiliation{Department of Applied Physics and Quantum-Phase Electronics Center (QPEC), University of Tokyo, Tokyo 113-8656, Japan}
\author{Y.~Kozuka}
\affiliation{Department of Applied Physics and Quantum-Phase Electronics Center (QPEC), University of Tokyo, Tokyo 113-8656, Japan}
\author{A.~Tsukazaki}
\affiliation{Department of Applied Physics and Quantum-Phase Electronics Center (QPEC), University of Tokyo, Tokyo 113-8656, Japan}
\affiliation{PRESTO, Japan Science and Technology Agency (JST), Tokyo 102-0075, Japan}
\author{M.~Onoda}
\affiliation{Department of Electrical and Electronic Engineering, Akita University, Akita 010-8502, Japan}
\author{H.~Aoki}
\affiliation{Department of Physics, University of Tokyo, Hongo, Tokyo 113-0033, Japan}
\author{M.~Kawasaki}
\affiliation{Correlated Electron Research Group (CERG) and Cross-correlated Materials Research Group (CMRG), RIKEN Advanced Science Institute, Wako 351-0198, Japan}
\affiliation{Department of Applied Physics and Quantum-Phase Electronics Center (QPEC), University of Tokyo, Tokyo 113-8656, Japan}
\affiliation{CREST, Japan Science and Technology Agency (JST), Tokyo 102-0075, Japan}

\today
\begin{abstract}
The sequence of prominent fractional quantum Hall states up to $\nu$~=~5/11 around $\nu$~=~1/2 in a high mobility two-dimensional electron system confined at oxide heterointerface (ZnO) is analyzed in terms of the composite fermion model. The temperature dependence of $\Rxx$ oscillations around $\nu$~=~1/2 yields an estimation of the composite fermion effective mass, which increases linearly with the magnetic field. This mass is of similar value to an enhanced electron effective mass, which in itself arises from strong electron interaction. The energy gaps of fractional states and the temperature dependence of $\Rxx$ at $\nu$~=~1/2 point to large residual interactions between composite fermions. 
\end{abstract}


\maketitle

\begin{table}[b]
\caption{\label{tab:table1} Comparison of basic characteristics for 2D charge carrier systems. Here, $\mb$ is the band mass, $\me$ the free electron mass, $\epsilon$ the dielectric constant, $\rs$ the Wigner-Seitz radius, $e^2/\epsilon \lB$ the Coulomb energy in Kelvin,  $\hbar\omega_\text{c}$ the cyclotron energy in Kelvin and $\zeta$ the Landau level mixing defined as the ratio of $e^2/\epsilon \lB$ to $\hbar\omega_\text{c}$. Magnetic field $B$ is given in Tesla and the charge carrier density $n$ in units of 10$^{11}$cm$^{-2}$.}
\begin{ruledtabular}
\begin{tabular}{ccccc}
 &\multicolumn{2}{c}{GaAs}&\multicolumn{1}{c}{SrTiO$_3$}&\multicolumn{1}{c}{ZnO}\\
  &electron&hole&electron&electron\\ \hline
 $\mb$/$\me$ & 0.069 & 0.38 & 1.26 & 0.29 \\
 $\epsilon$ & 13 & 13 & 20000 & 8.5\\
 $r_\text{s}$ & 1.8/$\sqrt{n}$ & 10/$\sqrt{n}$ & 0.02/$\sqrt{n}$ &  11.6/$\sqrt{n}$\\
 $e^2$/$\epsilon \lB$ & 50$\sqrt{B}$ & 50$\sqrt{B}$ & 0.03$\sqrt{B}$ &  75$\sqrt{B}$\\
 $\hbar\omega_\text{c}$& 19.4$B$ & 3.5$B$ & 1.1$B$ &  4.6$B$\\
 $\zeta$ & 2.6/$\sqrt{B}$ & 14.3/$\sqrt{B}$ & 0.027/$\sqrt{B}$ & 16.6/$\sqrt{B}$\\
\end{tabular}
\end{ruledtabular}
\end{table}

The emerging field of oxide heterostructures revealed exceptional physical characteristics at the interface between two oxide band insulators\cite{Hwang2004,IQHE_ZnO}, including superconductivity \cite{Caviglia2008,*Reyren2007}, the quantum Hall effect \cite{IQHE_ZnO, FQHE_ZnO}, magnetism \cite{Brinkman2007,*Yamada2011}, metal-insulator \cite{Thiel2006,*Cen2008} and insulator-superconductor \cite{Caviglia2008, *UenoSTO, *UenoKTO} transitions. Essential to the evolution of this field has been the material SrTiO$_3$ and its derived heterostructures, which has recently displayed an electron mobility exceeding 100,000 cm$^2$/Vs \cite{Stemmer2010}. However, in recent years, the MgZnO/ZnO heterostructure has emerged as an outstanding material for two-dimensional (2D) high-mobility electron system in this new realm of materials science, where steady improvements in growth technique have realized mobilities as high as 800,000~cm$^2$/Vs \cite{IQHE_ZnO, FQHE_ZnO, Falson2011}. In such a clean 2D system, the Coulomb interaction can dominate the physics in magnetic fields \cite{Tsui1982, CDW1979, CDW1996,*Lilly99, *MITKravchenko, *Wigner1934}. In low fields, interacting charge carriers can be treated within the Fermi liquid theory as non-interacting quasiparticles with renormalized effective mass and Land\'{e} $g$-factor\cite{Tan2005, *Pudalov2002, *Shashkin2003, *Shashkin2002,*Gokmen2008}. The interaction strength is characterized by the Wigner-Seitz radius ($\rs$) defined as a ratio of Coulomb energy ($\propto \frac{e^2}{\epsilon}\sqrt{n}$) to kinetic energy ($\propto \frac{n}{\mb}$), where $n$ is the charge carrier density, $\mb$ the band mass and $\epsilon$ the dielectric constant of the hosting material. In a high magnetic field $B$, the kinetic energy of charge carriers is quenched, where the Coulomb interaction ($\frac{e^2}{\epsilon\lB}$) on the magnetic length scale $\lB=\sqrt{\hbar/e B}$ leads to the fractional quantum Hall effect(FQHE). This sequence of FQHE is described by the composite fermion picture, which maps the original system in a magnetic field $B$ to a fermion system in an effective magnetic field $B_\text{eff}=B-2n\phi_\text{0}$  by attaching an even number of flux quanta $\phi_\text{0}$ to each electron\cite{perspective}. At $\nu=1/2$ in particular, the state is mapped to a CF system interacting with a Chern-Simons gauge field with a zero $B_\text{eff}$ at a mean-field level.  Conversely, beyond the mean field approximation, density fluctuations of CFs cause the fluctuation of effective magnetic field, and this dictates the inter-CF interaction, which is totally different from the original interaction at $B=0$.  The inter-CF interaction  also determines the renormalized mass $m_\text{CF}$ of a composite fermion\cite{onoda1,*onoda2}. 

Now the appearance of correlation effects, both in low and high magnetic fields, depends on the material parameters. Table~I compares them for ZnO-based 2D electron system (2DES) with the well-established GaAs and SrTiO$_3$ systems, which shows that large $\rs$ and strong Coulomb interaction in high magnetic fields distinguish ZnO from other materials. Therefore we anticipate the electron transport in this material to be strongly affected by the Coulomb interaction in weak as well as strong magnetic fields. Moreover, the electrons in ZnO occupy an isotropic single pocket around $\Gamma$-point in the Brillouin zone. This removes the effect of mixing of heavy and light charge carriers like in GaAs-based hole system and the effect of anisotropic mass like in AlAs and SiGe systems. Thus the novel oxide system is an appealing platform to study many-body effects in low dimensional system.

\begin{table}[t]
\caption{\label{tab:table2} Parameters for each sample: $\mu$ is the electron mobility in zero field, $n$ the charge carrier density, $r_\text{s}$ the Wigner-Seitz radius, $m$ the electron mass,  $\me$ the free electron mass, and $\tq$ the quantum scattering time.}
\begin{ruledtabular}
\begin{tabular}{cccdcd}
Sample&$\mu$(cm$^2$/Vs)&
\multicolumn{1}{c}{$n$(10$^{11}$cm$^{-2}$)}&
\multicolumn{1}{c}{$\rs$}&
\multicolumn{1}{c}{$m$/$\me$}&
\multicolumn{1}{c}{$\tq$(ps)}\\
\hline
A&770,000& 1.5 & 9.5 & 0.46 $\pm$ 0.03 & 2.2\\
B&550,000& 1.9 & 8.4 & 0.47 $\pm$ 0.03 & 3.1  \\
C&460,000& 1.8 & 8.7 & 0.47 $\pm$ 0.03 & 2.8 \\
\end{tabular}
\end{ruledtabular}
\end{table}

In this Letter, we study aspects of electron many-body effects in ZnO-heterostructures by investigating temperature-dependent magnetotransport, both in low magnetic fields and in the lowest 
Landau level, where a sequence of FQH states $\nu=\frac{p}{2p\pm 1}$ with $p$ up to 5 is observed, due to the significant improvement of electron mobility in the oxide structure~\cite{Falson2011}. This enables us to draw the comparison between the electrons and the composite fermions in ZnO-structures. 

Experiments are performed in three high-mobility heterostructures with the largest mobility $\mu$ up to 770,000~cm$^2/$Vs for a charge carrier density $n$~=~1.4$\times10^{11}$cm$^{-2}$. Differences between samples under study may be attributed to slightly varied Mg doping in the Mg$_x$Zn$_{1-x}$O/ZnO capping layer, as Falson \textit{et al.}, with the details of MBE growth, have reported elsewhere in detail\cite{Falson2011}. The zero-field mobility and the charge carrier density determined from Shubnikov-de-Haas~(SdH) oscillations are listed in Table~\ref{tab:table2}.  The Hall-bars are processed by the conventional optical lithography combined with Ar-ion milling to define the mesa and electron beam evaporation of 20~nm Ti to form the ohmic contacts. In the last processing step, the structure is covered with 30~nm thick amorphous Al$_2$O$_3$. The magnetotransport measurements are performed in a $^3$He system and in a magnetic field up to 14~T applied perpendicular to the 2DES plane. The 2DES is excited with 100~nA DC current while measuring $\Rxx$ and $R_\text{xy}$ simultaneously as a function of magnetic field. Heating effects caused by the excitation current were not encountered.

\begin{figure}[!thb]
\includegraphics{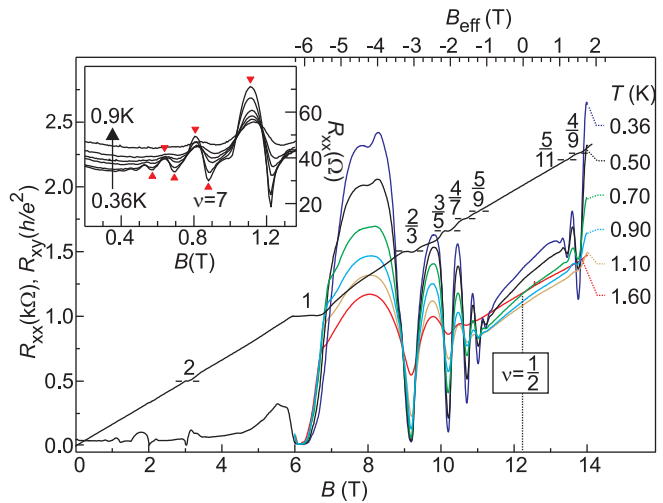}
\caption{\label{fig:1} (Color online) 
$\Rxx$ and $R_\text{xy}$ for sample A. In the lowest Landau level, a series of fractional quantum Hall states is observed. Temperature dependence of $R_\text{xx}$ at high magnetic field and in low field (inset) are shown. Triangles indicate the extrema of Shubnikov-de-Haas oscillations used to evaluate the electron effective mass.}
\end{figure}

Figure 1 shows the magnetotransport for the highest mobility sample A in magnetic fields up to 14~T. The high quality of the heterostructures is demonstrated by the unambiguous observation of a series of fractional quantum Hall states $\nu$ in the lowest Landau level at 
2/3, 3/5, 4/7, 5/9 for the fillings approaching $\nu=1/2$ from below, and 4/9, 5/11 from above.   The sample quality is further testified by 
indications, from $\Rxx$, of the formation of fractions down to $\nu$=6/11 and 6/13.  
We start by analyzing the temperature dependence of the amplitude $\mathit{\Delta} R$ of SdH oscillations in low magnetic fields using the Dingle expression $\mathit{\Delta} R \propto \exp(-\pi/\omega_\text{c}\tq)\xi/\sinh(\xi)$, where $\xi/\sinh(\xi)$ with $\xi=2\pi^2k_\text{B}T/\hbar\omega_\text{c}$ describing the temperature induced damping of SdH oscillations at cyclotron frequency $\omega_\text{c}=eB/m$. Here, we consider only the fundamental Fourier component of the magnetoresistance oscillation~\cite{Dingle1, *Dingle2}. An example of temperature dependence of SdH oscillations is depicted for sample A in inset of Fig.~1. The minima of SdH oscillations correspond to odd filling factors, which is consistent with previous reports, and are considered as electron cyclotron gaps\cite{Tsukazaki2008, *Kozuka2012}. For each value (indicated with a triangle) of the magnetic field, we analyzed $T$-dependence of $\ln(\mathit{\Delta} R/T)$to obtain an averaged electron mass $m~\approx~0.47\me$, where $\me$ is the free electron mass. Such analysis is performed for all three samples under study and the averaged electron masses are listed in Table~II.  Despite slightly different $\rs$ values for the samples, the electron mass enhancement is consistently high ($\sim 60\%$) above the band mass (0.29$\me$) of ZnO. Thus we conclude that the low-field mass enhancement is pronounced from electron-electron interaction, which saturates for such a large $\rs$ values.  The mass enhancement will lead to an enhancement of the LL mixing in FQH regime.  We assert our result by confirming that the quantum scattering time $\tq$, listed in Table~II, does not show a significant temperature dependence.  A large ratio of transport scattering time $\ttr$($\approx$100~ps) determined from the zero-field mobility to $\tq$ supports additionally the framework of large electron-electron scattering, and hence is consistent with the observation of an enhanced electron effective mass\cite{Gold1988}.

\begin{figure}[!th]
\includegraphics{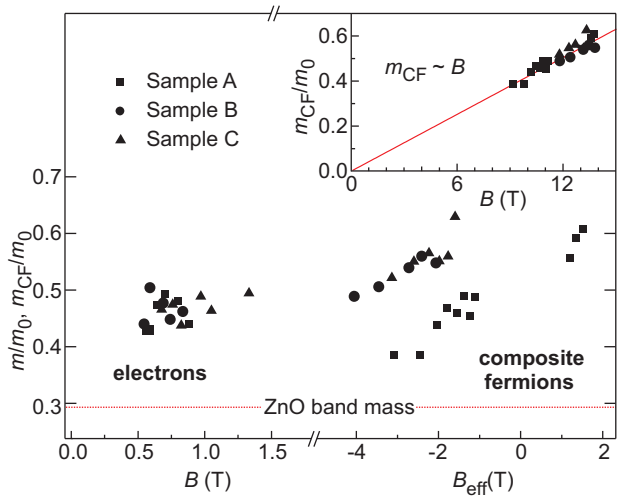}
\caption{\label{fig:2}  (Color online) The effective mass of electrons in low magnetic fields against  $B$, and the effective mass of composite fermions against $B_\text{eff}$. The mass in low magnetic fields is increased by about 60~$\%$ with respect to the electron band mass (horizontal line) 
in ZnO. Inset shows a linear dependence of the composite fermion effective mass $m_\text{CF}$ on the applied magnetic field $B$. The linear fit goes through the origin.}
\end{figure}

Now we come to the main question about the effective mass in the FQHE regime.  Treating the oscillations of $\Rxx$ in the lowest Landau level as SdH oscillations of CF, we consider the formation of a degenerated Fermi surface for CF at $B_\text{eff}$~=~0 and thus draw analogy between CF and electrons. Hence we apply the same Dingle analysis as described above to determine the effective mass of CF fermions in magnetic field $B_\text{eff}$ (upper axis in Fig.~1). From the temperature dependence of the $\Rxx$ extrema in Fig.~1, we evaluate the CF effective mass $m_\text{CF}$. The other two samples are analyzed in a similar manner and the summary of the derived CF masses is given on the right hand side of Fig.~2 as a function of $B_\text{eff}$.  Unlike in most observations for GaAs-based 2D systems, we do not observe the divergence of $m_\text{CF}$ when approaching $B_\text{eff}$~=~0, even though our data are analyzed in the same $B_\text{eff}$ range\cite{Du1994,*Shayegan1994,*Coleridge1995, *Leadley1994}. However, the data set of evaluated CF masses for three samples, when plotted versus the bare magnetic field $B$, fall upon a linear magnetic-field dependence as depicted in  inset of Fig.~2. This $B$ dependence of CF mass departs from the result of a dimensional analysis based on the CF mean field theory, which predicts $m_\text{CF}\propto\sqrt{B}/\epsilon$ scaling \cite{Halperin1993}. Beside the difference in functional dependence of $m_\text{CF}$ on the magnetic field, the CF mass in ZnO is also lower than the estimated value $m_\text{CF}$=0.9$\me$, when considering both $m_\text{CF}$ scaling with the dielectric constant $\epsilon$(see Table 1) and $m_\text{CF}$=0.65$\me$ for GaAs-based 2DES with comparable charge carrier density \cite{SmetCFDisperion}. However, the dimensional analysis does not include the effect of Landau level mixing, which is much larger in ZnO than in GaAs as displayed in Table~I, and the non-zero thickness of the wave function. Therefore, to interpret our result, the impact of these two effects on electron interaction in ZnO have to be taken into account\cite{ParkJain,Nayak2009,Yoshioka1984,*Yoshioka1986}.  The mass of electron, plotted on the left-hand side of Fig~2, is independent  of the magnetic field and is comparable with the CF mass. For comparison, the ratio of CF mass to the electron mass in GaAs is about 10. 

\begin{figure}[!th]
\includegraphics{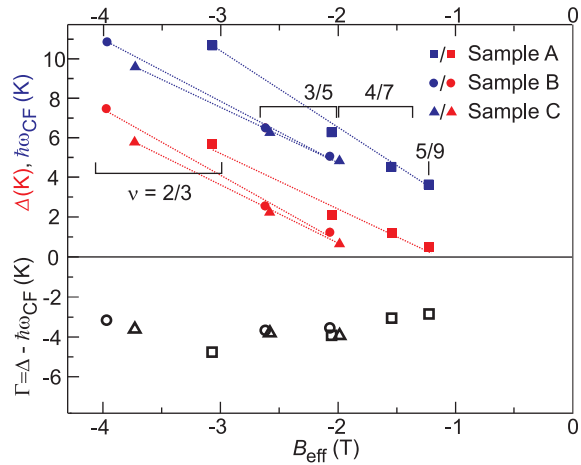}%CFMass.eps}
\caption{\label{fig:3}  (Color online) Comparison of the composite-fermion cyclotron energy $\hbar\omega_\text{CF}$ evaluated from Fig.~2, and the energy gap of quasiparticles $\mathit{\Delta}$ against $B_\text{eff}$. The dashed lines serve as guides to the eye. The open symbols in the negative-energy region represent the difference $\mathit{\Gamma}$ of energy gap $\mathit{\Delta}$ and cyclotron energy $\hbar\omega_\text{CF}$.}
\end{figure}

Having identified the CF mass, we next evaluate the energy gap as the CF cyclotron energy $\hbar\omega_\text{CF}$ for the identified fractional filling factors, and plot them in Fig.~3 (blue) against  $B_\text{eff}$. At $\nu$~=~2/3 the values of the cyclotron energies are $0.047\frac{e^2}{\epsilon \lB}$ and $0.042\frac{e^2}{\epsilon \lB}$ for sample A and B respectively, which are close to the activation energy of 0.1$\frac{e^2}{\epsilon \lB}$ ( 0.08$\frac{e^2}{\epsilon \lB}$) for spin-polarized (non-spin polarized) state of an ideal system \cite{perspective}. The values of the so-obtained cyclotron energy are larger than the energy gap $\mathit{\Delta}$ obtained from the temperature dependence of $\Rxx$ minima in terms of $\Rxx \propto \exp(-\mathit{\mathit{\Delta}}/2T)$. Figure~3 plots $\mathit{\Delta}$ (red) versus $B_\text{eff}$. As expected, the energy gap of fractional states decreases when $B_\text{eff}$~=~0 is approached, and reflects the tendency for the cyclotron energy $\hbar\omega_\text{CF}$ dependence on $B_\text{eff}$. The difference $\mathit{\Gamma}$ between the two energies, shown at the bottom of Fig.~3, does not indicate any pronounced dependence on the magnetic field. The difficulty to interpret the behavior of $\mathit{\Gamma}(B_\text{eff})$ arises from several factors. The heterostructures differ from each other by the Mg doping level, and even though its variation is small between the structures, the disordered potential landscape and the attributed correction $\mathit{\Gamma}_\text{disorder}$ should be different for three samples. A variation in Mg doping should also result in the variation of the conduction band offset between MgZnO and ZnO, which defines the 2DES confinement potential.  
%@Ultimately 
This will result in different thickness of the 2DES wavefunction in three heterostructures. However, it is difficult to estimate the variation of the confinement potential and the thickness of the wave function, given the current level of understanding of the system, namely a limited knowledge about the conduction-band offset between ZnO and the low level of Mg doping (roughly $x$ = 0.01) utilized to achieve high-mobility heterostructures. 

\begin{figure}[!th]
\includegraphics{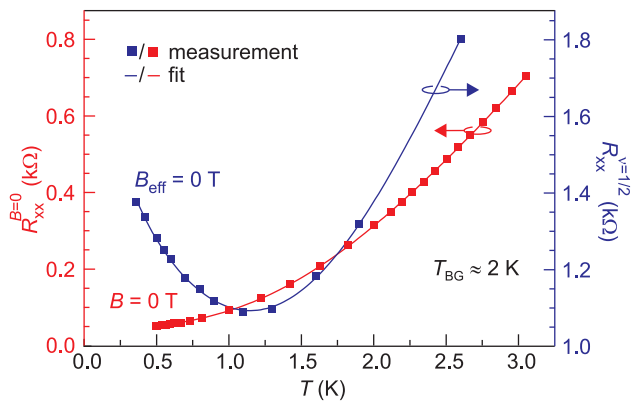}
\caption{\label{fig:4}  (Color online) Temperature dependence of $\Rxx$ at $B~=~0$~T (red)  and $\nu$~=~1/2 (blue) for sample A. Squares represent the measured data points, while lines are the best fits with Eq.~\ref{Rxx}. The Bloch-Gr\"{u}neisen temperature is the same for electrons and for composite fermions. }
\end{figure}

Finally, we compare the temperature dependence of $\Rxx$ for electrons at  $B$~=~0 and for composite fermions at $B_\text{eff}$~=~0. Figure~4 plots $\Rxx$ (filled symbols) for these values of magnetic field against temperature. We immediately notice a vast difference between the two.  
It turns out that we can fit the data with a phenomenological function, 
\begin{eqnarray}
R_\text{xx}=c_1+c_{2}T^2+c_3{\rm ln}T+c_{4}\frac{(T/T_\text{BG})^4}{1+(T/T_\text{BG})^3}, \label{Rxx}
\end{eqnarray}
where $c_1+c_2T^2$ describes the scattering of a degenerated Fermi system from crystal imperfections, while the logarithmic correction accounts for the interaction between the particles. The last term describes the phonon scattering in 2D systems with a characteristic Bloch-Gr\"{u}neisen temperature $T_\text{BG}=2k_\text{F}\hbar v_\text{s}/k_\text{B}$, which is related both to the sound velocity in the material $v_\text{s}$ and the Fermi wavevector $k_\text{F}$ \cite{Stormer1990,Kang1995}. The so-chosen phenomenological function reproduces rather accurately the temperature dependence at both magnetic fields, where a steeper drop of $\Rxx$ at $B_\text{eff}$~=~0 reflects a more efficient phonon scattering for CFs than for electrons, while $T_\text{BG}\approx$~2~K is the same for both carrier types. This temperature corresponds to $v_\text{s}$~=~1400~m/s in sample A, which is in the right range for sound velocity in ZnO and thus gives us a confidence in phenomenological function \cite{ZnOSound}. However, for CFs, with a $\sqrt{2}$-larger Fermi wavevector \cite{Halperin1993}, $T_\text{BG}$ is expected to be higher, when assuming a constant sound velocity $v_\text{s}$.  

More importantly, however, Eq.~1 reproduces the upturn of $\Rxx$ for CF given by the logarithmic term. The ln$T$ may remind us of the weak localization effect, but this is irrelevant, since $c_3$ is orders of magnitude larger at $B_\text{eff}$~=~0 than at $B=0$, and for high magnetic fields the physics will totally change. Rather, this should suggest a large residual interaction between the composite fermions [14].  The interaction, different from the bare Coulomb at $B=0$, comes from fluctuations in the Chern-Simons gauge field \cite{Halperin1993}. Then a marginal Fermi liquid behavior becomes possible in the presence of disorder, so we envisage  ln$T$  as arising from this kind of inter-CF interaction.  

To summarize, effects of the electron correlation have been analyzed in high-mobility oxide (ZnO) heterostructures for both weak and strong magnetic field regimes. In low magnetic fields, correlation manifests itself as a 60$\%$ enhancement of the electron mass compared to ZnO band mass. In high magnetic fields corresponding to filling factors around $\nu$~=~1/2, the data extending to fractional quantum Hall states $\nu=\frac{p}{2p\pm1}$ with $p$ up to 5 indicate that the ratio of CF mass estimated from $R_\text{xx}$ oscillations has a similar value to the enhanced electron mass, which contrasts with much higher values in GaAs.  Another contrast to GaAs is that the CF mass in ZnO increases linearly with the magnetic field and does not exhibit a divergence around $B_\text{eff}$=0. Finally, a comparison of the transport ($R_{xx}$) at $B=0$ and at $B_\text{eff}$=0 shows that the latter has a greater ${\rm ln}T$ component, indicative of a strong residual interaction between CFs.  CFs are also affected by phonon scattering.

Forthcoming studies will describe an investigation of the electron transport at lower temperatures and at high magnetic fields. This will give an insight into the effects of Landau level mixing and of the thickness of the electron wave function.  Of particular interest is the electron transport in high Landau levels, where the electron ground state is believed to form charge density waves originating from the competition between the electron correlation effects governed by Wigner-Seitz radius and the magnetic length \cite{CDW1996}.

We would like to acknowledge fruitful discussions with Kentaro Nomura, Koji Muraki and Lars Tiemann. This study is partially supported by the Japan Society for the Promotion of Science (JSPS) through the "Funding Program for World-Leading Innovative R$\&$D on Science and Technology (FIRST Program)", initiated by the Council for Science and Technology Policy (CSTP).

\end{document}